\documentclass[a4paper,twocolumn,superscriptaddress,showkeys,preprintnumbers]{revtex4-2}
\usepackage{amsmath}
\usepackage[hidelinks,urlcolor=blue]{hyperref}
\usepackage{notoccite}
\makeatletter
\renewcommand{\@biblabel}[1]{\quad#1.}
\makeatother

\usepackage{color}

\definecolor{Gray}{gray}{.25}

\usepackage{graphicx}
\graphicspath{figures}

\usepackage{sidecap}

\usepackage{wrapfig}
\usepackage[pscoord]{eso-pic}
\usepackage[fulladjust]{marginnote}
\reversemarginpar

\begin{document}

\title{Tuning the workfunction of ZnO through surface doping with Mn from first-principles simulations} 
		
\author{Andreas Douloumis}
\address{Dept. of Materials Science and Technology, University of Crete,  Heraklion, 70013, Greece}

\author{Nikolaos R.E. Vrithias}
\address{Dept. of Materials Science and Technology, University of Crete,  Heraklion, 70013, Greece}
\address{Institute for Electronic Structure and Laser, Foundation for Research and Technology - Hellas, Heraklion, 70013, Greece}

\author{Nikos Katsarakis}
\address{Department of Electrical and Computer Engineering Hellenic Mediterranean University (HMU), Heraklion, 71004, Greece}
\address{Institute for Electronic Structure and Laser, Foundation for Research and Technology - Hellas, Heraklion, 70013, Greece}

\author{Ioannis N. Remediakis}
\email{remed@materials.uoc.gr}
\address{Dept. of Materials Science and Technology, University of Crete,  Heraklion, 70013, Greece}
\address{Institute for Electronic Structure and Laser, Foundation for Research and Technology - Hellas, Heraklion, 70013, Greece}

\author{Georgios Kopidakis}
\address{Dept. of Materials Science and Technology, University of Crete,  Heraklion, 70013, Greece}
\address{Institute for Electronic Structure and Laser, Foundation for Research and Technology - Hellas, Heraklion, 70013, Greece}

\begin{abstract}
		
		Surface doping of ZnO allows for tailoring the surface chemistry of the material while preserving the superb electronic structure of the bulk. Apart from obvious changes in adsorption energies and activation energies for catalysis, surface doping can alter the workfunction of the material and allow it to be tuned for specific photocatalytic and optoelectronic applications. 
		We present first-principles electronic structure calculations for surface doping of Mn on the ZnO (0001) surface. Various dopant concentrations have been considered at the out-most (surface) layer of Zn atoms, while the interior of the material is kept at the ideal wurtzite structure. For each system, the surface energy and surface workfunction have been calculated.
		Both workfunction and surface energy drop with increasing Mn concentration for O-terminated surfaces, while more complex behaviour is observed in metal-terminated ones. We discuss trends in surface stability and surface electronic structure of this material and how they affect its properties.
\end{abstract}

\preprint{To be published in {\it Surface Science}}

	\maketitle

	\section{Introduction}
	ZnO has attracted a lot of attention as one of the best materials for light-matter interactions \cite{p1,p2}. Besides stability and abundance, ZnO exhibits an array of useful optical and electronic properties. It shows a wide band gap of about 3.3 eV which is direct, located at the $\Gamma$ point of the Brillouin zone, high electron mobility, transparency, and a large exciton binding energy of about 60 meV \cite{r6}. 
	ZnO can be produced by very easy crystal-growth techniques, allowing for low cost ZnO-based devices, it is not toxic for humans, has been used in the cosmetics and food industries for years \cite{r0}, and has been used in several technological applications such as sensing \cite{rem8}.
	Compared to narrow-band gap semiconductors such as Fe$_2$O$_3$, GaP and GaAs which are superb visible light absorbers \cite{sunlight,r4}, ZnO has excellent stability in aqueous suspensions and therefore it is suitable for photocatalytic applications 
	\cite{frr,rem9}.  
	ZnO can be found in many forms including wurtzite (B4), which is the most stable state at ambient conditions, as well as zinc-blende (B3) and rocksalt (B1) \cite{wurstable,r8}.
	
	Several methods have been employed in order to further improve the photocatalytic efficiency of ZnO, including dye sensitization, polymer modification, interface forming and others  \cite{dyes,r9,r44,r2,r16}. Perhaps the most popular method for tuning the properties of ZnO is doping \cite{r7,r3,dope}. While doping effectively reduces the wide bandgap of the material, it often creates unwanted states within the bandgap, modifying the good electronic properties of the photocatalyst and thus resulting in reduced efficiency \cite{unwanted,bad}.
	
	A common issue in photocatalysts is the unwanted fast recombination of electrons and holes, resulting to low efficiency. To address this, most photocatalysts include some form of a heterostructure, usually a metal-semiconductor or semiconductor-semiconductor one. A Schottky junction, formed when there is direct contact of a metal and a semiconductor, allows for only unidirectional electron flow facilitating the separation of photogenerated electron-hole pairs. Plasmons at the metal surface create intense and highly localized electric fields that drive electrons from the metal to the conduction band of the semiconductor. The energy of incident light needed to generate plasmons is less than the band gap of the semiconductor, so it is feasible to harvest the light energy of the visible part of the spectrum \cite{r32, hs}. On a carefully designed semiconductor-semiconductor junction, band offsets and Fermi level are designed so that electrons of the conduction band of one material  move to the valence band of the other, thus allowing for a separation of electron and holes \cite{sshs}.
	
	Workfunction is a key parameter that can be used to describe the properties of such heterostructures. For example, in the metal-semiconductor heterostructure the charge polarization formed is determined in large part by the difference between the workfunctions of the two materials. The workfunction is a good descriptor for the electron affinity of a facet and a highly effective descriptor for identifying the reduction and oxidation facets of a photocatalyst \cite{r50}. When placing two semiconductors adjacent to each other the one with the lower workfunction ought to have the higher Fermi energy, so knowing the workfunction allows for predictions on the effects of rational designs of these more complex photocatalytic structures. Thus, rational design of semiconductors requires an intimate understanding and tunability of the workfunction of the materials involved \cite{r17}.

	The importance of the workfunction and tuning thereof is well understood and highly sought after in optoelectronics and especially in the field of transparent conducting oxides, such as ZnO, where work function serves as one of the main descriptors governing the conducting behaviour \cite{henrichcox, mason2010,so2012}. As an example, in device applications relying on an Ohmic contact and charge transfer between the active channel layer and electrodes, even a small difference in the work functions of the different layers may result in a Schottky-like barrier. This may lead in reduced charge transfer between layers and thus reduced efficiency of the device \cite{lee2019}.  
	
	Many techniques have been developed to address the problem of fine-tuning the workfunction of a semiconductor. Sun et al. achieved that by creating an inteface of ZnO-MoO$_3$  \cite{r73}. Ghosh et al. \cite{r71} use the combination of a Au/ZnO interface and oxygen vacancies to that end. In the present work, we propose surface alloying as a method that can efficiently tune the workfunction while at the same time keeping intact the bulk electronic structure including direct badgap and high mobilities. In the following, we will demonstrate that surface doping can be used to change the workfunction of the material.

	As a typical example of a doped  oxide used in applications, we choose Mn-doped ZnO. This material
	has been studied extensively in the context of photocatalysis  \cite{rem10,rem6,rem7} . Apart from photocatalysis, this material has unique magnetic \cite{rem1, rem2}, optical \cite{rem3,rem11},  electrical \cite{rem4,rem5}, and sensing \cite{rem12,rem13} properties.  However, Mn doping often induces defects and alters the electronic structure of ZnO \cite{rem14}. Surface alloying, on the other hand, is expected to tune the surface properties without affecting the bulk. Although several works can be found in the literature studying bulk Mn-doped ZnO using first-principles simulations \cite{wang04,gallegos19}, only recently researchers have started simulating surfaces or thin films \cite{boukhari21}. Moreover, the electronic structure modification due to surface doping is not clear.	Here, we consider ZnO covered by a surface alloy with Mn, so that the out-most metal layer contains a mixture of Zn and Mn atoms. We calculate the workfunction of these surfaces for O-terminated, Zn-terminated surface, as well as a stoichiometric slab that has one O-terminated and one Zn-terminated surface.
	
In order to discuss the stability of surface doping, especially at high Mn concentrations, we calculate the surface energy of the systems. Surface energy is the energy cost per unit area of creating a surface; a negative value implies that the material would disintegrate \cite{lodziana04}. Surface energy for a composite system, such as a doped oxide surface, depends on the total energy of the system, which is readily available from the {\it ab initio} calculations, and also on the chemical potentials of the components \cite{kolasinski,kolasinski2}. We calculate the surface energy of ZnO for various values of the chemical potentials, and then for the Mn-doped ZnO. We use these results to comment on the stability of each surface.
		
	\begin{figure}
		\includegraphics[width=\linewidth]{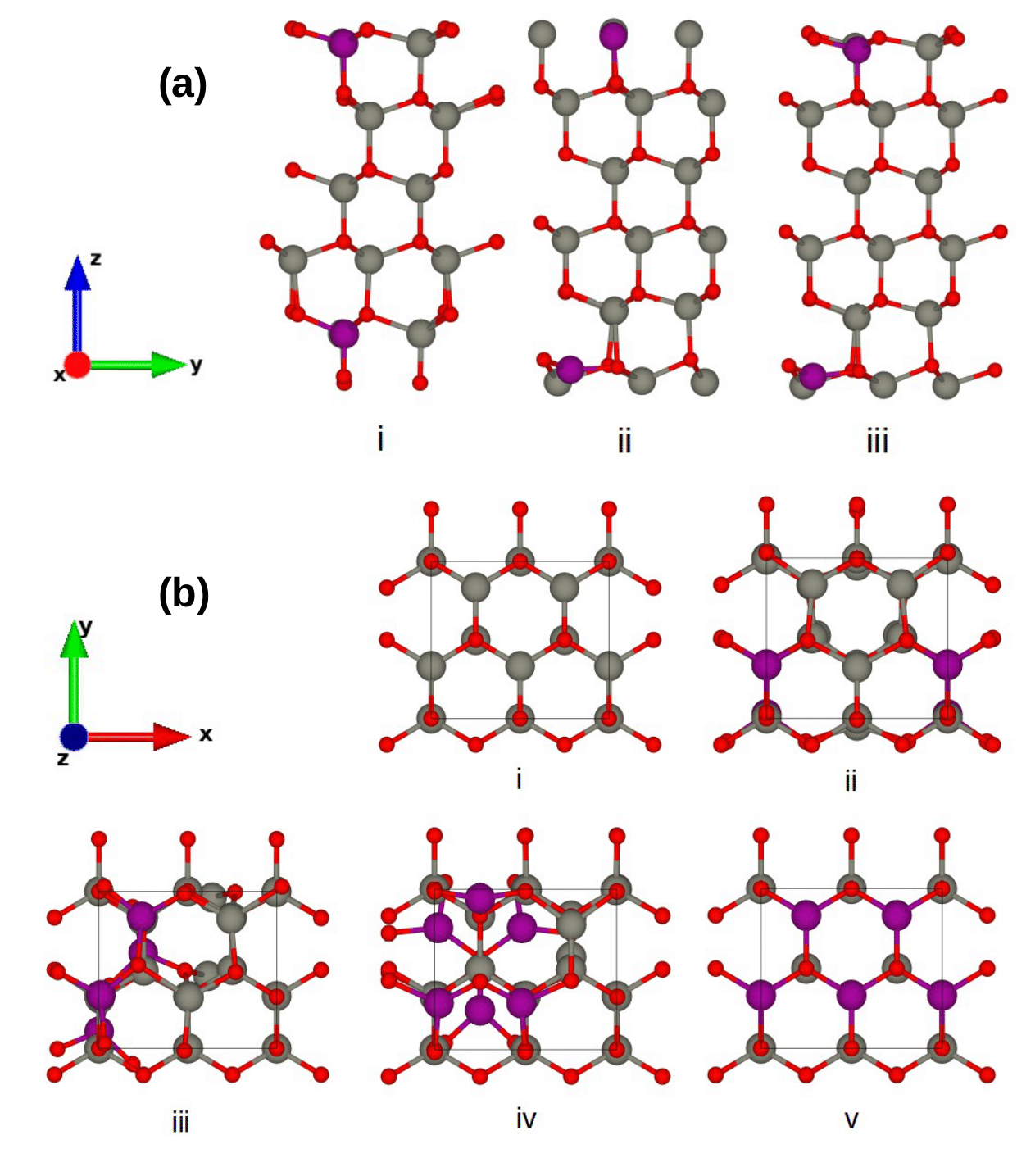}
		\caption{Ball-and-stick models of representative configurations used in the present study (a): Side view of relaxed ZnO slabs with different terminations at 25\% Mn surface doping. From left to right: (i) Oxygen-terminated, (ii) Metal-terminated, (iii)  stoichiometric. (b): Top view of relaxed stoichiometric ZnO slabs with increasing \% concentration of Mn on the surface double layer:  (i) 0\%, (ii) 25\%, (iii) 50\%, (iv) 75\%, (v) 100\%. Zinc, Oxygen and Manganese are shown in gray, red and purple respectively.  }
		\label{fig1}
	\end{figure}

	\section{Computational Details}

	We model ZnO(0001) using slabs with $(2\times 2)$ periodicity. Each slab consists of six ZnO layers, while each (double) layer contains one layer of four O atoms and one layer of four metal atoms. Periodic boundary conditions are imposed along all axes and a vacuum of 24 \AA\  is used along the $z$ directions.
	We consider three different terminations for ZnO as is shown in Fig.\ref{fig1} (a) to account for both polar and non-polar surfaces. In particular, the ideal stoichiometric slab contains 24 metal atoms and 24 O atoms and has one O-terminated and one metal-terminated surface. The metal-terminated slab is obtained by removing four O atoms from the O-terminated side and the oxygen-terminated slab is created by removing four Zn atoms from the metal-terminated side. 
	
	For each one of these three slabs, we replace Zn atoms of the out-most layer on each side of the slab by Mn.  Although the concentration of Mn in bulk Mn-doped ZnO is of the order of few percent, it is possible that much higher concentrations of Mn might decorate the surfaces  \cite{rem13,rem14}. To this end, we consider ZnO surfaces with high Mn content on the surface layer, while there is no Mn in the subsurface layers. As each layer contains four Zn atoms, we can create  Mn$_x$Zn$_{1-x}$ layers with $x = 0.00, 0.25, 0.50, 0.75$ and $1.00$. In total, we consider fifteen different systems (three different terminations combined to five different Mn concentrations). Some examples of these configurations are shown in Fig. \ref{fig1} (b). 
	We have tried several other values for the number of layers in the slab in order to make sure that our simulation is converged with respect to the slab thickness.

	First-principles calculations were performed using density functional theory (DFT). 
	The Vienna Ab-initio Software Package (VASP) software in the Projector-Augmented Wave (PAW) mode was used for the entirety of these calculations \cite{vasp1, vasp2, vasp3, vasp4}. 
	For the exchange correlation functional we implement the generalized gradient approximation (GGA) by Perdew-Burke-Ernzerhof (PBE) \cite{r49}. A  Mohnkrost-Pack set of $3\times 3\times 1$ ${\mathbf k}$-points was used to sample the Brillouin Zone. A plane-wave basis with cutoff energy of 500 eV was used. For the slab calculations, a dipole correction was employed.	We made several test runs with the DFT+U method as well as spin-polarized DFT and a hybrid exchange-correlation functional. While these methods give better description of the electronic structure of bulk ZnO, and more accurate values for the band gap, they increase computational time significantly, while they yield similar values for the workfunction as the plain DFT. For example, using DFT+U with a value of 6 eV \cite{huang2012} for the U parameter, the workfunction of a metal-terminated ZnO slab is found to be 4.25 eV, while DFT gives  4.29 eV, so that both values round up to 4.3 eV. For this reason, we use standard DFT throughout this paper.

	For slab calculations, the simulation box, and therefore the wurtzite lattice constants, were kept frozen to the values we found for bulk. We find $a=3.28$ \AA{} and $c=5.33$ \AA{} for the lattice constants of wurtzite ZnO, in very good agreement to experimental values of $a=3.25$ \AA{}  and $c= 5.20$ \AA\  \cite{explc}.
	The atomic positions of two (double) layers on each side of the slab are allowed to relax during the simulation. The two (double) layers in the middle of the slab are kept intact to the positions of bulk ZnO.  ZnO surfaces, and in particular O-terminated ones, have complex reconstruction, that can affect the work function \cite{pal13}. As we are only interested in trends in workfunction with Mn doping, we only consider few structures for each Mn concentration. Although we limit our simulation to the $(4\times 4)$ cell, we have tried several different initial configurations, which were subsequently fully relaxed, and we choose the minimum-energy structure among several candidate structures for each Mn-content and surface termination.

	We use the  VESTA package \cite{vesta} for visualization of atomic structures (Figs. \ref{fig1} and \ref{wfplot}) and the  Gnuplot package \cite{gnuplot} for graphs (Figs. \ref{wfplot},\ref{wf},\ref{pretty},\ref{se}).

	\section{Results and Discussion}

	\subsection{Workfunction}
	
\begin{figure}
		\includegraphics[width=\linewidth]{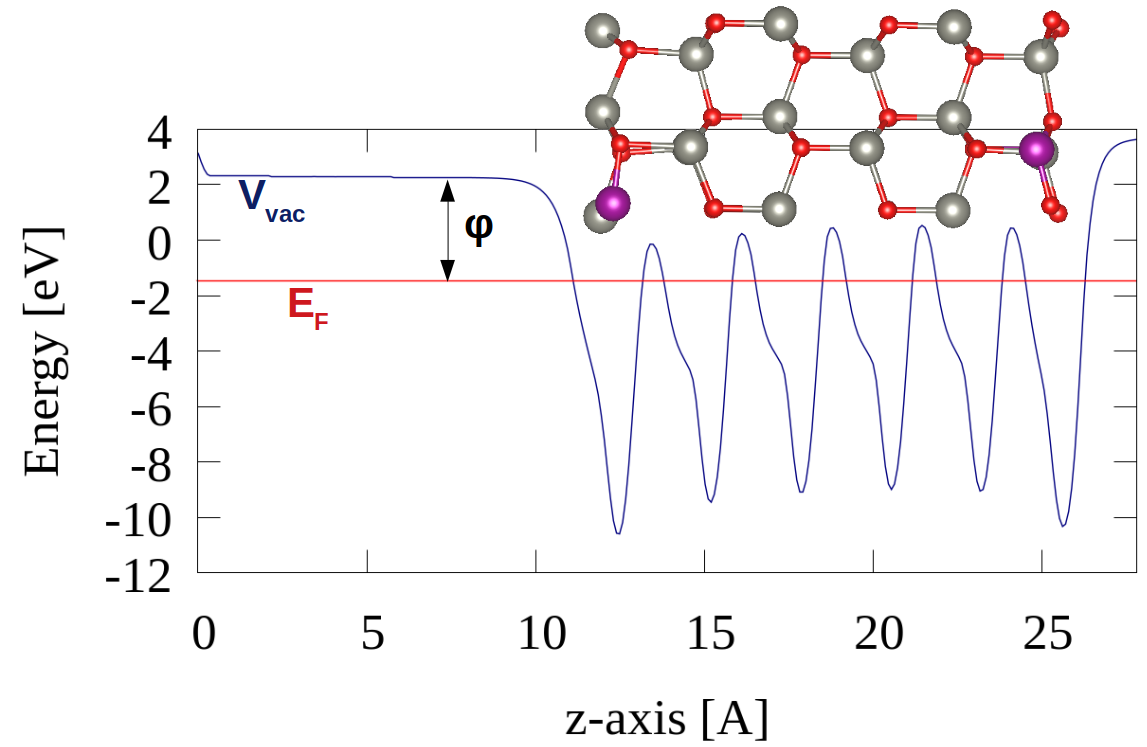}
		\caption{Local part of the average total electrostatic potential of electrons along the $z$ axis for a stoichiometric slab with 25\% Mn. The vacuum potential $V_{vac}$, Fermi energy $E_F$ and workfunction $\phi$ are also shown, as well as a ball-and-stick model of the structure (Mn: purple; Zn: gray; O:red). }
		\label{wfplot}
	\end{figure}

	The workfunction ($\phi$) is defined as the minimum energy needed to transfer an electron from the Fermi energy ($E_F$) to the Vacuum level ($V_{vac}$) \cite{kittel}: 
	
	\begin{equation}
		\phi=V_{vac}-E_F
		\label{eq:vvac}
	\end{equation}
	In order to calculate $V_{vac}$, we plot the local part of the total electrostatic potential of electrons as a function of distance from the surface. In the vacuum region, away from the slab, the local potential $V$ forms a plateau where it takes its vacuum value $V_{vac}$. A typical plot that is used to determine $V_{vac}$ is shown in Fig. \ref{wfplot}.
	
	Calculating $V_{vac}$ for a complex polar material, in particular one without mirror plane in the middle of the slab, is a non-trivial task. In most cases, the charge transfer induced by Mn, causes the potential to have large oscillations in the vacuum region. Also, the asymmetric slab together with the imposed periodic boundary conditions, result in a non-zero electric field in the vacuum region. As a result, the potential in the vacuum region is not constant, but rather proportional to distance, as is the case for the potential energy inside a constant electric field. A typical remedy for these problems is to include an artificial dipole layer in the slab, so that the electric field in the vacuum region is zero. We have used this so-called dipole correction in all calculations. Even then, the parameters of the dipole layer must be carefully tuned in order to obtain flat potential energy.
	Moreover, the exchange-correlation potential introduces large numerical errors in regions of zero electron density. For this reason, the potential energy used in this work does not contain terms that should be zero in the vacuum region, namely the exchange-correlation potential and the non-local part of the Kohn-Sham Hamiltonian.  
	
	 For each system, we have made several calculations with different dipole parameters or different parameters for the electronic relaxation, in order to obtain flat potential energy in the vacuum region like the one shown in Fig. \ref{wfplot}. The values of workfunction we report here are average values of workfunctions from the two sides of the slab.

\begin{figure}
		\includegraphics[width=\linewidth]{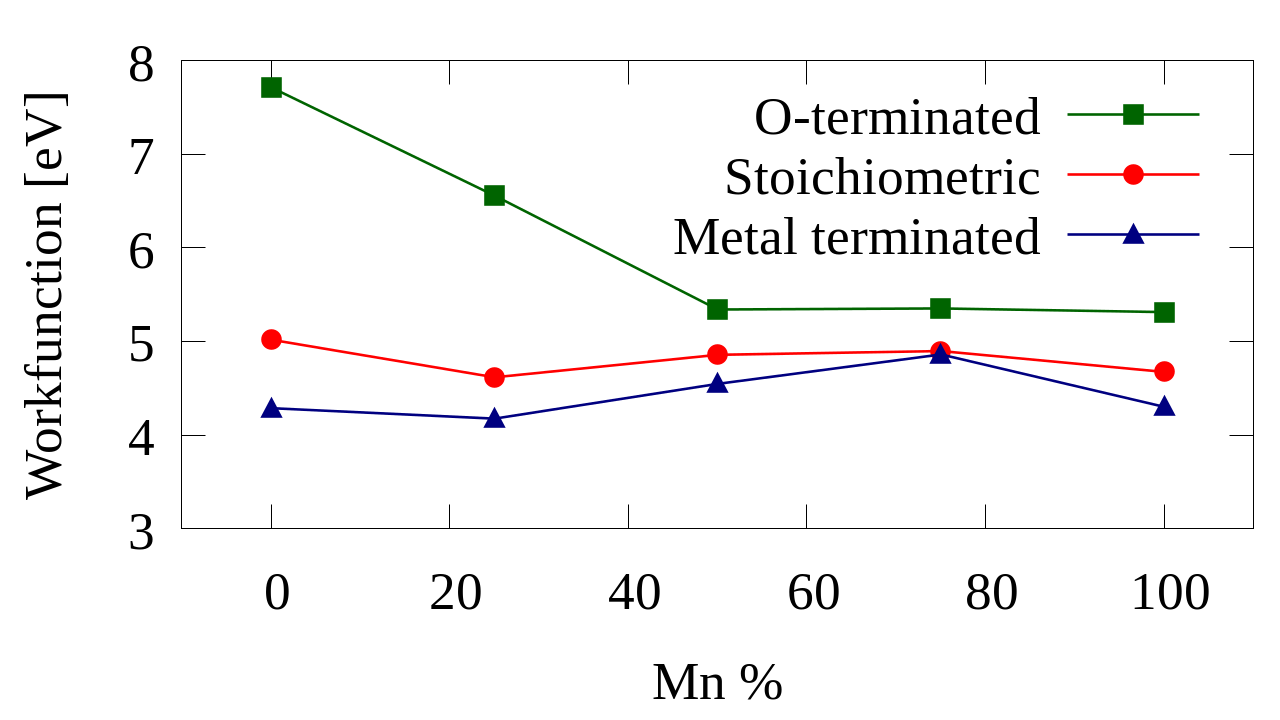} 
		\caption {The Workfunction $\phi$ as a function of the surface concentration of Manganese for the different terminated configurations.}
		\label{wf}
	\end{figure}

	The calculated workfunctions for all different terminations and Mn \% on the surface can seen in Fig. \ref{wf}. Changes on the surface concentration of Mn lead in changes in the workfunction of about 0.5-0.7 eV for the metal-terminated and stoichiometric cases while for the oxygen-terminated case the effective range of the workfunction is almost 3 eV wide.

	The work function is higher at the O-terminated surface than the metal-terminated surface. The stoichiometric slab has one O-terminated side and one metal-terminated side and has values that lie between the values for the O-terminated and the metal-terminated slab. For the stoichiometric slab of pure ZnO with zero Mn, we find workfucntion to be $\phi= 5.0$ eV, in agreement with $\phi= 5.0$ eV in the calculation of Ghosh et al. \cite{r71} and very close to the experimental value of 4.7 eV \cite{rem15}.  For the O-terminated slab, our value of $\phi= 7.7$ eV is close to the value of 6.9 eV calculated by Sun et al. \cite{r73}.

	Surface alloying results in reducing the work function at small Mn concentration. A possible explanation could be related to the fact that Mn is slightly less electronegative than Zn. Pauling electronegativities are 1.65 for Zn and 1.55 for Mn. As a result, the Mn-O bond is slightly more polar than the Zn-O bond and therefore electrons are slightly more bound to Mn-doped ZnO. For higher Mn concentrations, the distortion caused to the material due to the presence of the surface alloy might lead to weakened chemical bonds and less stable electrons. This could be the case for the local maximum of workfunction around 75\% concentration of Mn on the metal-terminated slab. Such non-monotonous behaviour of the workfunction of ZnO with increasing alloying is also observed for MoO$_3$ overlayers on ZnO \cite{r73}.
	
	Introducing Mn on the surface can result to vast changes on the workfunction, by almost 3 eV for the case of O-terminated surface. For comparison, Au and O defects can change the workfucntion by about 1 eV \cite{r71}. This fact could be used for applications where large changes of the workfunction are desirable, the most prominent one being ZnO interfaces used in photocatalysis. 	
		
	Such strong dependence of the workfunction on surface composition might cause some surprise. From the definition of Eq. (\ref{eq:vvac}), the workfunction seems to depend on bulk properties, while surface alloying only affects the surface of the material. The reason that doping alters the workfunction is the same reason that causes different $(hkl)$ faces of a crystalline material to have different workfunctions: The abrupt termination of a material at a surface causes some of its electrons to be localized near the surface, creating a quasi-metallic layer near the surface. This layer will create a jump in the electrostatic potential between the material and the vacuum, as shown at the right end of the plot in Fig. \ref{wfplot}. The amount of surface dipole depends on the detailed atomic structure of the out-most layer, leading to the variation in work function for different surfaces \cite{bardeen}.
	
	Those surface states are widely known for pinning the Fermi level, making methods like doping ineffective in altering the workfunction of the material in applications such as the Schottky junction. The surface alloying technique completely alters the surface chemistry of the material, and thus the surface states. So while there is a change in the local Fermi energy level close to the surface, the bulk energy levels, as well as the rest of the bulk properties of the material remain intact.
	
		\begin{figure}
		\begin{center}
			\includegraphics[width=\linewidth]{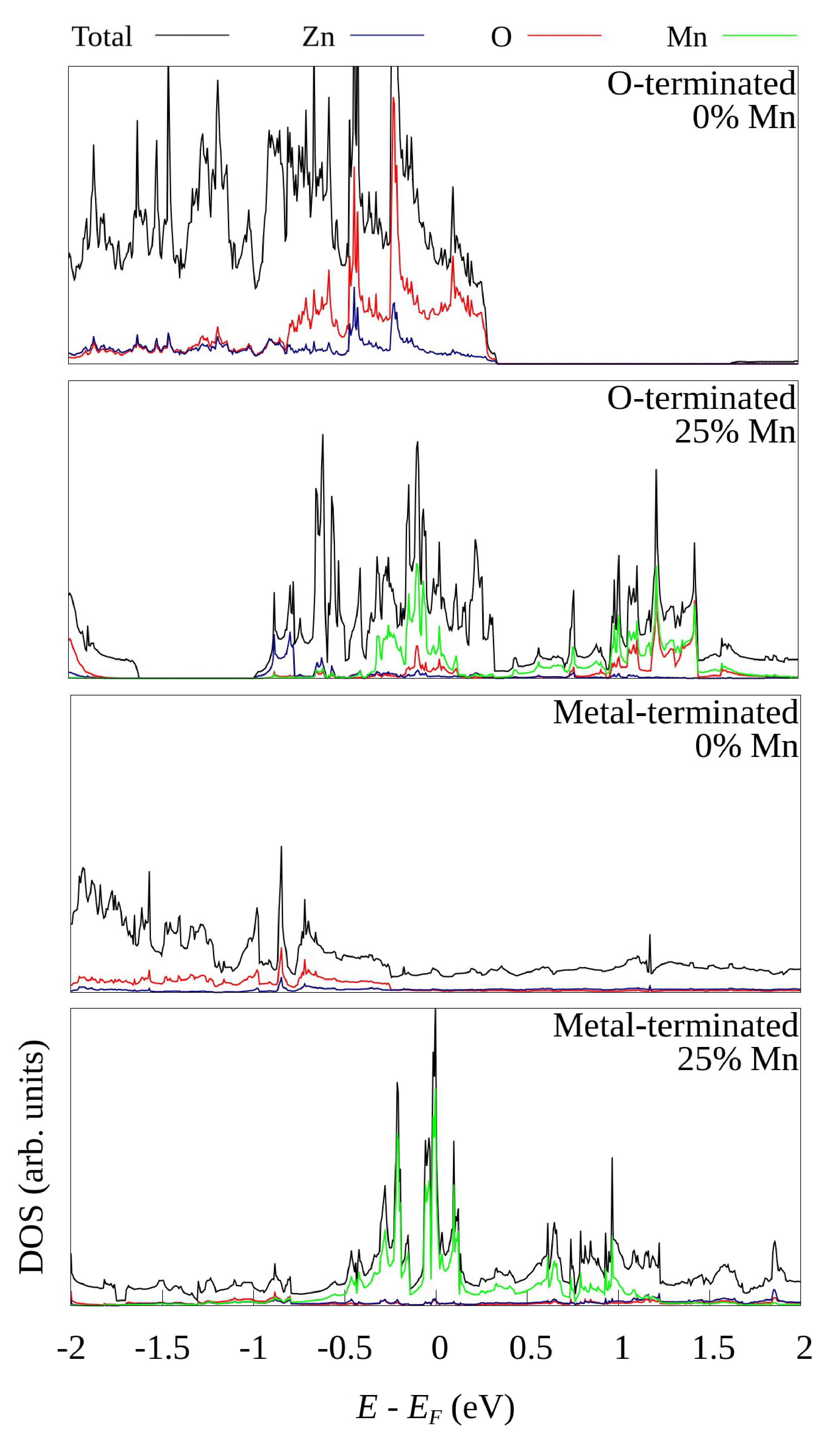} 
		\end{center}
		\caption {Total- and atom-projected Density-of-states (DOS) plots for O-terminated and metal-terminated ZnO slabs with and without Mn doping at the out-most layers. The O, Mn and and Zn atoms used for the projection of the DOS are those of the out-most (surface) layers.}
		\label{figdos}
	\end{figure}
	
    The results for the workfunction shown in Fig. \ref{wf} show that doping creates a much stronger effect for the O-terminated slab compared to the stoichiometric and the metal-terminated slabs. To address this interesting finding, we look at the atom-projected densities of states (DOS) for O-terminated and metal-terminated slabs with 0 and 25\% Mn at each out-most surface layer. The results are shown in Fig. \ref{figdos}. For Mn-free surface, the electrons close to the Fermi level originate from O atoms. In particular, the O-terminated surfaces have electron-rich states manifested by several high peaks around the Fermi level in the DOS plot. These peaks coincide with the atom-project DOS for the O atoms of the out-most layers of the O-terminated slab. On the contrary, the metal-terminated slab has mush fewer electrons around the Fermi level, and the DOS diagram is almost flat around zero. Although many different atom types contribute to the DOS around the Fermi level for the metal-terminated slab, the largest contribution comes from O atoms that are below the Zn layer. 
    
    Upon introduction of Mn, the DOS plot changes dramatically, with new peaks around the Fermi level. Naturally, these new peaks originate from the Mn atoms, and the Mn orbitals dominate the electronic structure near the Fermi level. So the electrons that determine the Fermi level, and consequently the workfunction, are mostly Mn ones. In order to accommodate all electrons of the system, higher states have to be occupied which results in an increase of the Fermi level of the surface. This increase of local Fermi level results to a lowering of the workfunction, see Eq. (\ref{eq:vvac}).  A rough descriptor for the increase of Fermi level could be the integral of the DOS from -1.5 eV to 0 eV.  The choice of the lower bound of the integral is of course arbitrary; we choose -1.5 so that the whole Mn peak is included. For the O-terminated slabs, the integral of DOS from -1.5 to 0 eV drops by about 60\% upon introduction of Mn. On the other hand, a drop of only 10\% is found for the metal-terminated slabs. This could explain why the workfunction changes more dramatically for the O-terminated slab compared to the metal-terminated one.

	\subsection{Surface Energy}
	
Surface alloying may lead to thermodynamic instabilities close to the surface, due to distortions introduced to the lattice by the different atom size of the alloying material. The standard descriptor of thermodynamic stability is the surface energy, defined as the excess energy of a slab compared to the same amount of bulk material, divided by the total area:   

%
	\begin{equation}
		\gamma = \frac{E_{slab}-\sum_i N_i \mu_i }{2A}
		\label{eq:gamma}
	\end{equation}
where $E_{slab}$ is the total energy of the slab, index $i$ runs over all atom kinds of the system (in our case Zn, O and Mn), $N_i$ is the number of atoms of kind $i$, and $\mu_i$ is the chemical potential of atoms of kind $i$. 
	
    A common use of Eq. (\ref{eq:gamma}) in surface chemistry: to a first approximation, the surface energy changes by $E_a/A$, where $E_a$ is the adsorption energy \cite{barmparis1,barmparis2}. Here, we use this equation to account for non-stoichiometric and alloyed systems, where the bulk energy depends on the bulk form of the components. The chemical potential, $\mu_i$, equals the energy per atom in a reservoir of $i$ atoms that is at equilibrium with the system under study.
    
    For Mn, we take the chemical potential to be equal to the energy per atom in bulk Mn. At standard conditions, Mn forms two complex crystal structures, known as $\alpha$- and $\beta$-Mn which are both cubic and contain several tens of atoms per unit cell. At elevated temperatures, Mn transforms first to the fcc structure, known as  $\gamma$-Mn, and then to the bcc structure, known as $\delta$-Mn \cite{wyckoff}. We calculated both the fcc and bcc structures of Mn and found lattice constants that minimize the energy. We find very small difference of 60 meV per atom between the two structures. We use the energy of bcc Mn as the chemical potential of Mn in Eq. (\ref{eq:gamma}).
    
     For Zn and O, the chemical potential depends on the alloying conditions one wants to simulate. We consider all possible combinations between the most stable structures for Zn, O  and ZnO. We calculated the energy for O$_2$ molecule, bulk wurtzite ZnO and bulk metallic Zn. Following the ideas of the grand-canonical ensemble of statistical mechanics, we consider our system to be in equilibrium with reservoirs of Zn and O atoms, that are characterized by the respective chemical potentials, $\mu_{Zn}$ and $\mu_{O}$. Depending on experimental conditions, $\mu_{Zn}$ will have value somewhere between the energies of a Zn atom in bulk Zn and the energy difference between ZnO and O$_2$ molecule. Similarly, $\mu_{O}$ will have value somewhere between the energies of an O atom in O$_2$ molecule and the energy difference between ZnO and bulk Zn. To include all possible reservoirs, we  introduce two parameters, $x$ and $y$ so that the chemical potentials are given by the following equations:
		
	\begin{equation}
		\mu_{Zn} = xE_{Zn} + (1-x)(E_{ZnO}-{\frac{1}{2}E_{O_2}}), 
		\label{mu1}
	\end{equation}
\begin{equation}
		\mu_{O} = y \frac{1}{2}E_{O_2} + (1-y)(E_{ZnO}-E_{Zn}),
		\label{mu2}
	\end{equation}

\begin{figure}
	\centering
	\includegraphics[width=\linewidth]{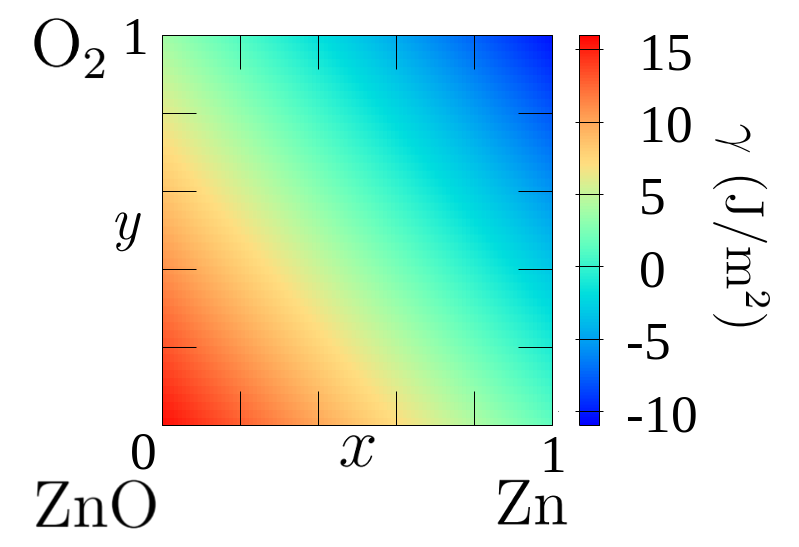} 
	\caption {Contour plot of the surface energy, $\gamma$, of a metal-terminated ZnO slab as a function of Zn and O chemical potentials. The chemical potential of Zn ranges between the values for bulk ZnO and bulk Zn ($x$ parameter, Eq. (\ref{mu1}) ). The chemical potential of O ranges between the values for bulk ZnO and O$_2$ molecule ($y$ parameter, Eq. (\ref{mu2}) ). }
	\label{pretty}
\end{figure}
	
	In Fig. \ref{pretty} we plot the surface energy of metal-terminated ZnO as a function of $x$ and $y$. For small values of $x$ and $y$, roughly in the range $x+y < 1$, the surface energy is positive, while it becomes negative for values of $x$ and $y$ that are close to 1.  The lowest energy of the system is when the O atoms that are removed from the slab to make it metal-terminated form O$_2$ molecules ($y=1$). Naturally, O$_2$ is by far the preferred thermodynamic state of oxygen. Another way to create metal-terminated slab is by adding Zn atoms to it. Now we need to subtract atoms from the reservoir, therefore the lowest value of $\gamma$ happens when Zn is at its least favourable state, which is bulk Zn ($x=1$). A Zn atom is more stable at ZnO than at bulk Zn, as $E_{ZnO}-{\frac{1}{2}E_{O_2}} < E_{Zn}$. This inequality comes from the fact that ZnO has negative standard enthalpy of formation, $\Delta H = E_{ZnO}-{\frac{1}{2}E_{O_2}} - E_{Zn}$, a fact that we also verify by direct calculation that yields $\Delta H = $ -2.8 eV, identical to the value calculated by Meyer \cite{meyer04}. 
	
	Values of surface energies as low as $-10$ J/m$^2$ or as high as $+15$ J/m$^2$ might be unrealistic, as most experimentally reported surface energies are of the order of few J/m$^2$. So the experimentally relevant conditions should be close to the diagonal of the square corresponding to $x+y \approx 1$. We used the full range of chemical potentials when creating Fig. \ref{pretty} just to give an idea of the extrema of the system. In the following, we use the more realistic values $x=0$ and $y=1$ for the analysis of the Mn-doped slabs.
	
		\begin{figure}
		\includegraphics[width=\linewidth]{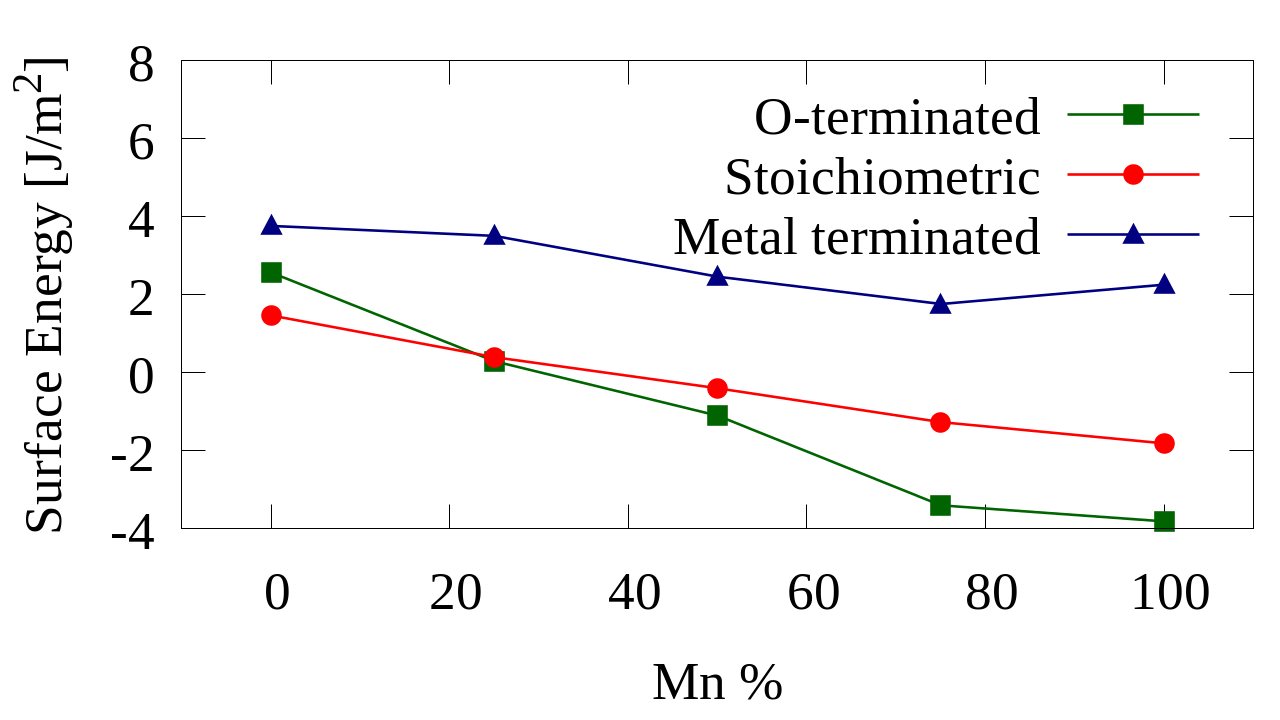} 
		\caption {The surface energy, $\gamma$, as a function of the concentration of Mn in the out-most metal layers for O-terminated, metal-terminated and stoichiometric ZnO slabs.}
		\label{se}
	\end{figure}

	For the Mn-alloyed surfaces, we consider only the case of O$_2$ molecules and ZnO, which corresponds to $x=0$ and $y=1$ in Eqs. (\ref{mu1}) and (\ref{mu2}) or the upper left corner of Fig. \ref{pretty}.  Results for the surface energy of the system as a function of Mn concentration on the out-most metal layers are shown in Fig. \ref{se}. Introducing Mn lowers the surface energy. This is a direct consequence of the much stronger bonds formed by Mn compared to bonds of Zn. This is shown in the large difference between the cohesive energy of the elemental solids, which are 1.35 eV/atom for Zn and 2.92 eV/atom for Mn \cite{kittel}. The gain by the strong Mn bonds is more important than any strain introduced to the lattice due to the 
	different sizes of atoms. For comparison, the radii of Mn and Zn in the elemental solids are 1.26 \AA\ and 1.39 \AA, respectively \cite{kittel}. This large difference of atom sizes leads to a small increase of surface energy for the metal-terminated slabs and for 100\% Mn compared to the 75\% Mn system. When the out-most layer consists of only Mn atoms, the symmetry is broken and the Mn atoms form two different bonds. This behaviour is not observed in the O-terminated surfaces where the surface energy drops monotonically with increasing Mn content. 
	  
	O-terminated slabs have negative surface energy for high Mn concentrations. This implies that these surface alloys might not be thermodynamically stable, as further cleavage is favoured. Although O-terminated and stoichiometric slabs are energetically unfavourable at high Mn contents of the outmost layers, they are found to be metastable states of the system. Including results for these systems provides extra evidence that both the workfunction and the surface energy of O-temrinated ZnO(0001) drops with increasing surface Mn, all the the way up to full Mn coverage. 
	
	As it was the case with workfunction, the surface energy of ZnO slab drops upon Mn doping, and the drop is more pronounced for the O-terminated slab. From the DOS plots of Fig. \ref{figdos}, it is clear that a large portion of the Mn-induced states lie above the Fermi level and they are therefore anti-bonding. As such, they will weaken the chemical bonds of the surface, making it easier to cleave the material, and thus making the surface energy to become lower. This effect is stronger at the O-terminated slab where Mn destroys the electron-rich states near the Fermi level, as shown in Fig. \ref{figdos}.

\begin{figure}
		\includegraphics[width=\linewidth]{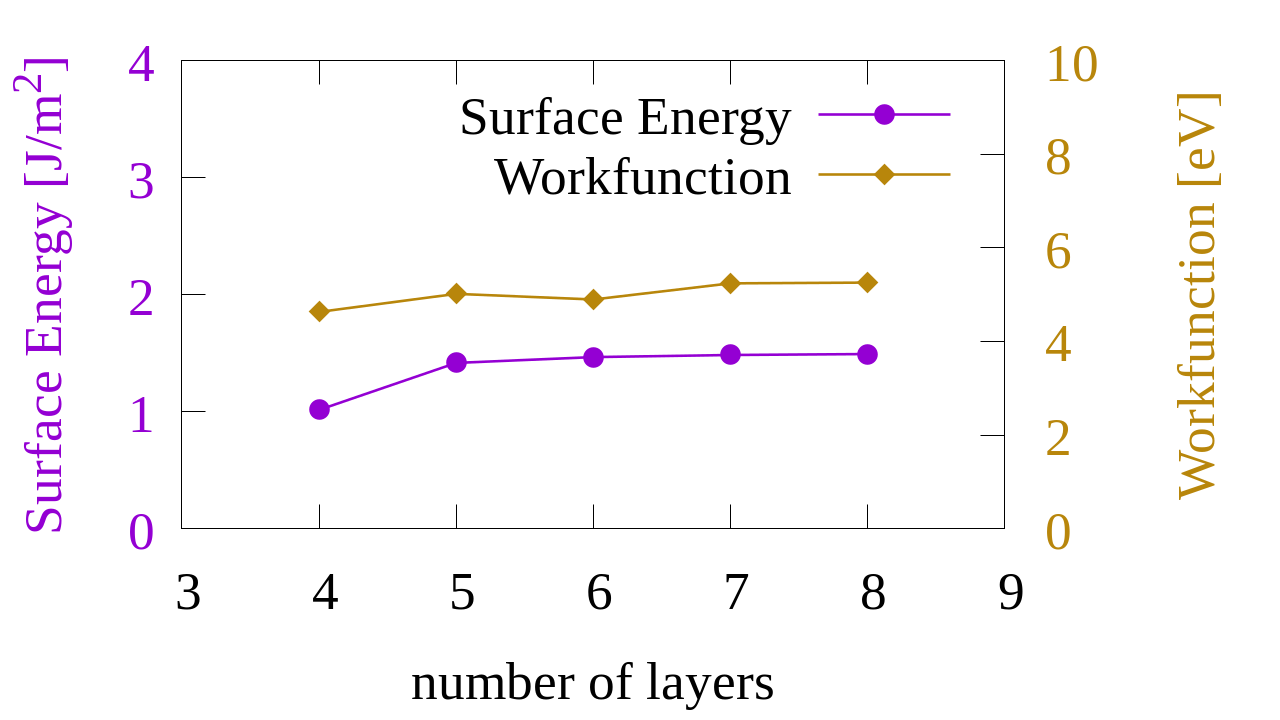} 
		\caption {Workfunction (left vertical axis) and surface energy (right vertical axis) for stoichiometric ZnO(0001) slab as a function of number of ZnO bilayers in the simulation cell.}
		\label{se+wf}
	\end{figure}

	As the surfaces we consider are strongly polar ones \cite{tasker1979,noguera2000}, it could be possible that the dipole moment induced by the dipole correction might be insufficient to overcome artefacts related to the finite  size of the system. This could cause oscillations in the calculated properties depending on whether there are odd or even number of layers in the simulation cell. For this reason, we calculate the work function and surface energy for various slab thicknesses in order to make sure that our results are converged. Fig. \ref{se+wf} shows calculated work function and surface energy as a function of the number of layers. The results seem to have small odd-even oscillations. To a good approximation, results do not depend on the number of layers after a thickness of five layers. 
	
	\section{Conclusions}
	f
	We studied the effects of surface doping of ZnO with Mn by means of DFT simulations of workfunction and surface energy. While there are few Mn atoms overall, and the material can be refered to as Mn-doped, all Mn is on the topmost surface layer, resulting to a surface alloy. This material has essentially the same atomic and electronic structure as ZnO, having however very different surface properties. Our results suggest that fine-tuning of the workfunction using this method of surface alloying is feasible and competitive compared with other options. The workfunction can be modified by several eV's. However, introducing large concentrations of Mn on the topmost metal layer of O-terminated ZnO results in thermodynamically unstable system. On the other hand, metal-terminated ZnO is stable under all Mn concentrations on its surface layer.
	
	\medskip
	\textbf{Acknowledgements} \par 
	We acknowledge useful discussions with Dr. George Kenanakis. 
	This work was supported by the Research Committee of the Univeristy of Crete (Grant KA10136) as well as by the Hellenic Foundation for Research and Innovation through project MULTIGOLD, grant HFRI-FM17-1303 /  KA10480. We acknowledge 
	computational time granted 	from the National Infrastructures for Research and Technology
	S.A. (GRNET S.A.) in the National HPC facility, ARIS, under
	projects pr007027-NANOGOLD and pa181005-NANOCOMPDESIGN.
	
	\medskip
	\textbf{Conﬂict of Interest}
	The authors declare no conﬂict of interest.
	
	\medskip
	\textbf{Data Availability Statement}
	The data that support the findings of this study are available from the
	corresponding author upon reasonable request.

\end{document}